# A New Charging Method for Li-ion Batteries: Dependence of the charging time on the Direction of an Additional Oscillating Field

I. Abou Hamad[a], M. A. Novotny[b,c], D. O. Wipf[d], P. A. Rikvold[a]

[a] Department of Physics, Florida State University, Tallahassee, Florida 32310, USA
[b] Department of Physics & Astronomy, Mississippi State University, Mississippi State, Mississippi 39762, USA
[c] HPC$^2$ Center for Computational Sciences, Mississippi State University, Mississippi State, Mississippi 39762, USA
[d] Department of Chemistry, Mississippi State University, Mississippi State, Mississippi 39762, USA

We have recently proposed a new method for charging Li-ion batteries based on large-scale molecular dynamics studies (I. Abou Hamad *et al*, *Phys. Chem. Chem. Phys.*, **12**, 2740 (2010)). Applying an additional oscillating electric field in the direction perpendicular to the graphite sheets of the anode showed an exponential decrease in charging time with increasing amplitude of the applied oscillating field. Here we present new results exploring the effect on the charging time of changing the orientation of the oscillating field. Results for oscillating fields in three orthogonal directions are compared.

## Introduction

Lithium-ion batteries with liquid electrolytes are widely used, from portable electronics to electric vehicles. We have used the revolutionary development of computational algorithms and computer hardware to simulate the anode half-cell of a Li-ion battery and propose a new charging method based on the application of an oscillating field in addition to the constant charging field (1). In that study, the orientation of the oscillating electric field was perpendicular to the graphitic sheets of the anode.

In this paper, we further investigate the effect of the orientation of the electric field on the Li-ion intercalation time or charging time, using molecular dynamics simulations. Results from simulations with the electric field oriented in two orthogonal directions parallel to the sheets are compared with those for the perpendicular direction. Moreover, we include preliminary results showing the organization of the electrolyte molecules in the double layer and conjecture a possible mechanism for the penetration of the Li-ion through the double layer and into the electrode.

## Model System and Simulation Method

As in the initial study (1), the model system is composed of graphitic sheets stacked to represent the anode, an electrolyte of ethylene carbonate (EC) and propylene carbonate (PC) molecules, hexaflourophosphate counter ions, and Li-ions (Figure 1).

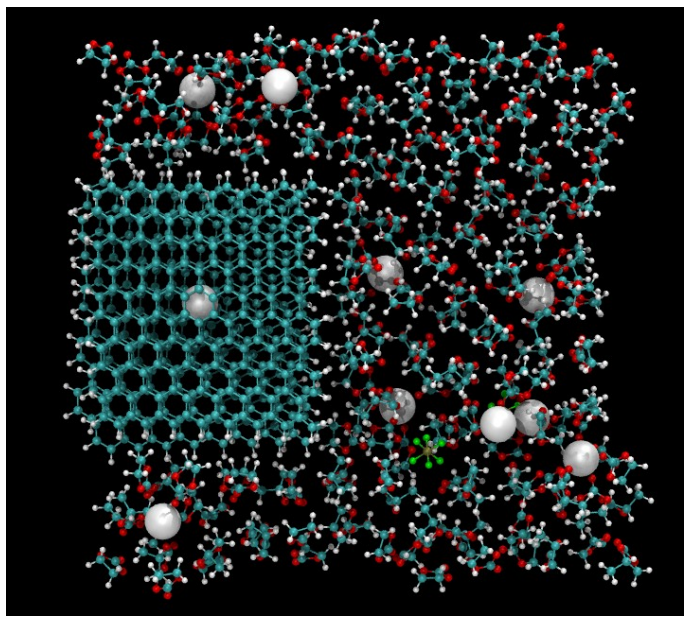

Figure 1. Model system after one of the Li-ions has intercalated. The white spheres represent the Li-ions, ethylene carbonate and propylene carbonate are the electrolyte molecules and the green ions are the hexaflourophosphate ions. Top view showing the plane of the graphitic sheets where the Li-ion is intercalated.

The general amber force field (GAFF) (2) was used for the bonded and van der Waals' interactions while the point charges on all atoms were approximated at the Hartree-Fock/6-31g* level using the Spartan simulation package (Wave-function, Inc., Irvine, CA). The molecular dynamics simulation package NAMD (3) was used for running the simulations and the visualization package VMD (4) was used for visualization and data analysis. For further detail we refer the reader to the original study (1).

To simulate a constant charging field, the charges on the carbons of the graphitic sheets were set to -0.0125 $e$ per atom. Moreover, the graphitic sheets were fixed from one edge to keep them stacked in a pillar formation. We have previously shown that applying an additional oscillating field in a direction perpendicular to the plane of the graphitic sheets ($Z$ direction) allows for much faster intercalation of the Li-ions (1). Here we compare the results to simulations with the oscillating field oriented in the other two orthogonal directions ($X$ and $Y$, parallel to the edges of the graphite sheets). Since faster intercalation is allowed with a larger amplitude of the oscillating field (1), all the simulations are run with an amplitude $A = 6.5$ kCal/mol and a frequency $f = 25$ GHz. Simulations were run in the NVT ensemble after the volume had been equilibrated in the NPT ensemble at constant pressure of 1 atm and temperate 300 °K. For each orientation of the oscillating field, ten simulations were performed.

**Simulation Results**

Intercalation Time

The intercalation is considered to be a Poisson process (5), thus the fraction of simulations in which Li-ions have not intercalated by time *t* for a given orientation of the oscillating field is

$$P_{non} = e^{(-t/\tau)}, \qquad [1]$$

where $\tau$ is the average interaction time for that process.

Figure 2 shows the fraction of non-intercalated ions as a function of time, as well as an exponential fit of the form in equation [1] to the data points. The average intercalation time for each direction of the oscillating field is also shown.

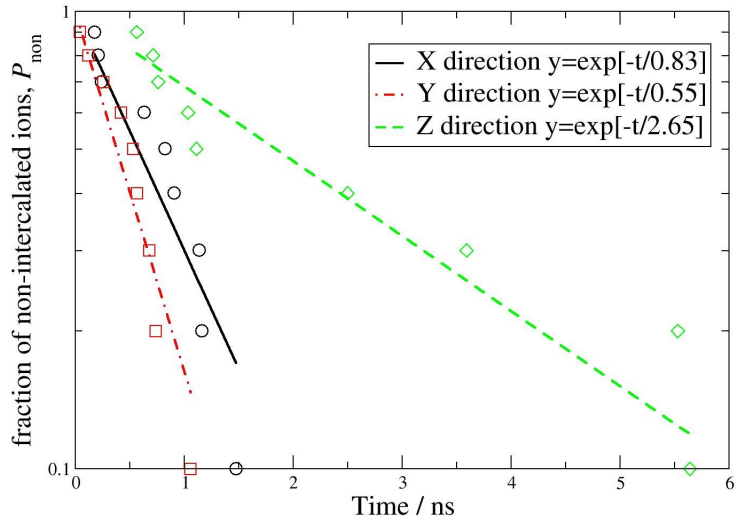

Figure 2. Fraction of non-intercalated Li-ions *versus* time. The lines are fits to the form given in equation [1]. The fitted parameters for each orientation of the oscillating electric field are given in the legend. *Z* is the direction perpendicular to the plane of the graphitic sheets. The average intercalation time is the denominator of the exponential.

The average intercalation times shown in figure 2 confirm the results we have shown in (1) and show that the results are not due to the orientation that was chosen. On the contrary, the average intercalation times for the other two orientations (*X* and *Y*) are shorter than for the *Z* orientation. This behavior should offer clues to the intercalation mechanism.

## Summary

We have investigated different orientations of the additional oscillating electric field and shown that similar results are obtained for the orthogonal directions to the one perpendicular to the planes of the graphitic sheets chosen in reference (1). Moreover, the intercalation times we obtain using those orthogonal directions are shorter than the one previously obtained in ref. (1).

## Acknowledgments

This material is based upon work supported by the Advancement of Hybrid Portable Power Systems under Contract No. W15P7T-09-C-S628 and by NSF grant No. DMR-0802288. Any opinions, findings and conclusions or recommendations expressed in this material are those of the authors and do not necessarily reflect the views of the CECOM Contracting Center.